\documentclass[fleqn,twoside]{article}
\usepackage{amsmath,amssymb,espcrc2}

\input BoxedEPS
\SetRokickiEPSFSpecial  
\HideDisplacementBoxes

\newcommand{\negsp}{\vspace*{-\bigskipamount}}
\newcommand{\fract}[2]{{\textstyle\frac{#1}{#2}}}

\let\lsim\lesssim
\let\gsim\gtrsim
\let\overdot\dot

\begin{document}

\title{Inventory and Outlook of High Energy Physics}
\author{Frank Wilczek\\
\small\it Center for Theoretical Physics\\ 
\small\it Massachusetts Institute of Technology\\ 
\small\it Cambridge, MA 02139-4307\\ 
\small MIT-CTP \# 3829 \\[1ex]
 Summary talk at ICHEP 2002, Amsterdam, July 2002.\\
\footnotesize I have kept very
close to the content and style of the talk as it was delivered. You   may access the
associated PowerPoint presentation through a link at
http://www.ichep02.nl/MainPages/PlenaryProgram.html\\[-1pc]}

\maketitle

\markboth{Frank Wilczek}{Inventory and Outlook of High Energy Physics}

%
\thispagestyle{empty}
 
\noindent
John Wheeler's description of the progression from graduate student to professor
as being the passage from knowing everything about nothing to knowing nothing about
everything carries over with obvious (and inessential) modifications to the
progression from talks at parallel sessions to the conference summary.    I feel
especially daunted here, after listening to a series of beautiful plenary talks that
have already summarized their subject areas, so that I'm reduced to providing a
summary of the second order.    In order to supply at least a small amount of value
added, I'll try to frame the new developments actually reported at the Conference
within the broader context of contemporary physics, and also to indicate a few
directions that I feel are important and promising, but which did not get
emphasized during the Conference.  
\vspace*{-\bigskipamount}

\section{The Standard Model}

\subsection{QCD}
\noindent
Very near half of the parallel sessions, 23 out of 48, centered around the strong
interaction and QCD.    This concentration reflects an enormous wealth of intriguing
phenomena.   We've been shown that even questions that have been pursued for
a long time, ranging from low-energy spectroscopy and the search for exotics to
high-energy reaction dynamics and the emerging new phenomenology of
diffractive dissociation at high energy, are far from being exhausted.   Still,
progress in the mature areas of strong interaction physics tends to be
incremental, and its description is necessarily complex, so I won't devote anything 
like equal time to reviewing them.   Instead, I'll content myself with a few general
observations about QCD.

The discovery that the basic degrees of freedom in the strong interaction  (quarks
and gluons) are entirely different from the apparent ones (mesons and baryons),
and the development of techniques, both theoretical and experimental, to work
with these degrees of freedom, is a wonderful triumph of empirical investigation
and intellectual analysis -- one of the greatest ever, I think.   Concealed beneath a
complex and bewildering array of phenomena, profoundly simple and
mathematically precise principles of  symmetry (nonabelian gauge invariance) and
dynamics (local quantum field theory) are found to be running the show.   

\begin{figure}[htb]
\centerline{\BoxedEPSF{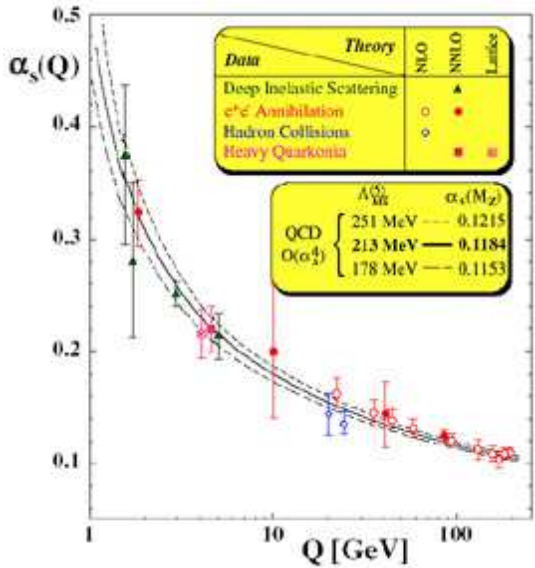 scaled 1500}}
\negsp
\caption{Running of the coupling in QCD.  It exhibits the uniformly successful
application of a theory containing only a single continuous parameter to describe a
tremendous variety of experimental measurements performed over a wide range of energies. 
Many of the calculations must be carried to several orders of perturbation theory to achieve
adequate accuracy.  Especially noteworthy is the ``lattice gauge theory'' point, which is
fully non-perturbative.}\label{FWf1}
\negsp
\end{figure}

No brief summary can do justice to this story, but most of you are familiar with its
main outlines, and Figure~\ref{FWf1} \cite{bethke} will serve both as icon and mnemonic.   In it
the results of dozens of experiments of different characters performed at a variety
of mass scales, together comprising many thousands of independent
measurements, are compared with the theoretical calculation of the running
coupling, which contains exactly one adjustable parameter.   The quality of the
agreement speaks quite well for itself, of course, but two amplifications are
appropriate.  

First, you will notice that the alternative theoretical curves, based on different
values of the adjustable parameter, focus to the right, at high energy, where most
of the data is.    The parameter is usually taken to be
$\Lambda_{\textrm{QCD}}$, roughly speaking the mass or inverse distance scale at
which the QCD coupling becomes of order unity.  On very general heuristic
grounds we should expect $\Lambda_{\textrm{QCD}}$ to be associated with
characteristic physical phenomena like the inverse of the geometric radius of
hadrons, as defined by high-energy total cross-sections, or the transverse
momentum `cutoff' observed for particle production in high-energy scattering.  
The equivalent energy scales are a little fuzzy to define, but they surely lie in the
range 150--250~MeV.    Not coincidentally, these values parameterize the limiting
curves in
Figure~\ref{FWf1},
which bracket the data.    The point I want to reinforce is that
if we regard this range of values for
$\Lambda_{\textrm{QCD}}$ as given, the theoretical predictions at high energy are
essentially unique.  They contain \emph{no adjustable parameters whatsoever!}    

Second, I want to call your attention especially to the theoretical point labelled
``heavy quarkonia-lattice''.   All the other theoretical points are based on
calculations using what are usually called ``perturbative methods''.  Actually, these
calculations usually employ sophisticated renormalization group
arguments and factorization techniques, besides including at least one and usually
two orders in virtual loops, so the name hardly does them justice.   In any case the
``perturbative'' calculations in themselves represent a {\it tour de force\/} of
quantum field theory, and they provide overwhelming, tangible evidence for the
quark-gluon description of strong interactions in general and the existence of jets
-- tangible incarnations of the quarks and gluons, which quite literally track
their energy-momentum --  in particular.   But the lattice calculations go yet
deeper.   They are based directly on the fundamental degrees of freedom and
algorithmic definition of the theory, with no approximations at all aside from
practical ones imposed by finite computer resources.   By way of contrast, the
corresponding calculations in QED, or even standard model electroweak theory, are
impossible in principle -- those theories lack nonperturbative definition, and
don't provide adequate foundations for completely honest calculations!  So it is
very reassuring, and gratifying, that the lattice calculations agree with the
perturbative ones, and with experiment.  Indeed the power of full,
nonperturbative implementation of the theory to determine
$\Lambda_{\textrm{QCD}}$ quantitatively already competes with that of traditional
perturbative methods, and it waxes steadily.

While stringent testing of the principles of QCD remains an important activity, the
main focus of research in the field has long since moved from verification to
application.   An ironic consequence of asymptotic freedom is that extreme
conditions of matter become amenable to reasonably straightforward theoretical
analysis, while great ingenuity (or extremes of patience and computer power) is
needed to use the theory in tamer circumstances.   Thus first-principles QCD
enables us to make accurate predictions for jets at large transverse momentum in
ultra-high energy collisions, providing the essential foundation for designing
signatures and estimating backgrounds for experiments at the energy frontier,
while ordinary nuclear physics remains mostly out of reach.   

On the other hand, recently there has been great progress in studying what might
be called extreme nuclear physics -- understanding the behavior of hadronic
matter in highly excited states \cite{alford} \cite{RW}.  

One limit of great interest is that of high temperature $T$, with zero net baryon
number density, or small chemical potential $\mu$.  According to theory, this is
the form of matter that dominated the Universe during the earliest moments of
the Big Bang.   Intuitively we might expect that energetic quarks, antiquarks, and
gluons are liberated, and propagate quasi-freely, at asymptotically high
temperatures.   And this is basically what happens; but to leave it at that would
constitute criminal neglect.   The high $T$, small $\mu$ regime of QCD is susceptible
to direct simulation using Monte Carlo techniques.   From this work it emerges that
the transition from the energy density characteristic of a pion gas, with 3 degrees
of freedom, to nearly that of free quarks and gluons -- 52 degrees of freedom! --
occurs in a narrow range of temperatures around $T_c \sim$ 150--200 MeV.  There
is no true phase transition, but a rapid crossover from a state of matter that would
be considered ``obviously hadronic'' to one best described starting with quasi-free
quarks and gluons, i.e., the famous quark-gluon plasma.   Although it has been
known for several years, I still find this precocious approach to the quark-gluon
plasma amazing, since on the low-temperature side one is not much past a dilute
gas.   On the other hand the pressure lags somewhat behind, and approaches
closely to the energy density only at $T\sim$ 600--800 MeV (whereas of course for
free massless particles they'd be equal).    A long-standing fundamental challenge
has been to understand the origin of this discrepancy analytically; it now appears
that this challenge is being met \cite{schroder}. 

Experimental work in heavy ion collisions is going great guns, as we've heard \cite{harris}.   
There is good evidence for elliptic flow,  which can be interpreted as a
manifestation of pressure, which implies multiple scattering.   When this is
analyzed quantitatively, it provides pretty convincing circumstantial evidence
that something approaching thermal equilibrium is established, at a temperature
well above the predicted crossover.   An important goal for the near future is to
gather direct evidence for ultra-high temperatures in the initial fireball, by
capturing prompt photons and dileptons.    Perhaps the most dramatic
experimental discovery so far in this field is the opaqueness of the fireball.  This
leaves its signature in monojets: high transverse momentum jets lacking a
balancing jet in the opposite hemisphere.    The opacity seems too large to be
ascribed to quasi-free quarks and gluons, and may be telling us something
important about the effective degrees of freedom activated in hadronic
wavefunctions.   It will be interesting to see how the result reported in this paragraph gets
reconciled with the result reported in preceding one!

Another lively frontier is the opposite limit, large $\mu$ and small $T$.  This is
relevant to the description of ultra-dense cold matter, as might be found in the
interior of compact stars.   By adapting techniques from the theory of
superconductivity we can construct a weak-coupling but nonperturbative
description of the ground state and low-energy excitations of three-quark matter
at asymptotically high densities.  In this remarkable color-flavor locked state
confinement and spontaneous chiral symmetry breaking, dynamical properties of
QCD that are often regarded as difficult and mysterious can be derived
rigorously and understood simply.    At sub-asymptotic densities we lose rigorous
control, and several more complicated alternatives have been proposed.   These
include interesting inhomogeneous and quasi-crystalline states, some
incorporating meson condensates.     Unfortunately both laboratory and numerical
experiments in this domain are impractical.   Compact stars do not reveal their
inner secrets easily, but technique in experimental astrophysics is in a state of
perpetual revolution, and we might be able to discern the effects of exotic states in
their mass-radius relations, cooling curves, and/or the gravitational radiation
they emit at creation or in violent collisions.     

\subsection{Electroweak}

\noindent
Although not traditional, I think it is useful and natural to discuss the electroweak
sector of the standard model as two separate theories, which are on a very
different conceptual footing and only loosely connected.   Here I refer not to the
two gauge symmetries $SU(2)\times U(1)$, which are birds of a feather, but rather
to two parts of the theory that really are radically different, namely the piece
describing gauge field interactions and the piece describing Higgs field
interactions.   (Note that every interaction of the standard model involves one or
the other!) 

\begin{figure}[htb]
\centerline{\BoxedEPSF{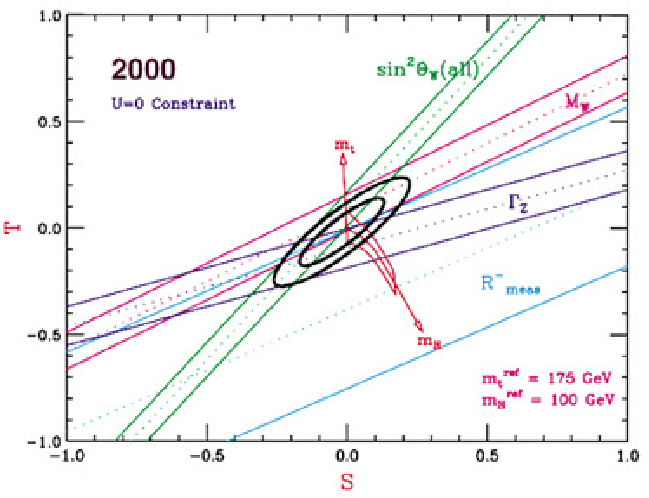 scaled 1000}}
\negsp
\caption{Comparison of precision electroweak data with experiment.  The accuracy
of the measurements is such that effects of virtual particles in loops must be carefully included. 
Indeed, these kinds of comparisons allowed the mass of the top quark to be
estimated before its discovery, and permit us to estimate the value of the Higgs
particle.  At present there is no convincing discrepancy between accurate
experiments and standard model predictions.}\label{FWf2}
\negsp
\end{figure}

The interactions of the gauge bosons are governed by a very tight theory, derived
from the profound and beautiful principle of local symmetry.   All the properties of
the gauge bosons, and thus of the interactions they mediate, must be derived
from just three continuous parameters:
$g_2$ and $g_1$, and the magnitude of the symmetry-breaking condensate, which
generates their mass.   (To be completely correct I should qualify this.   
Within the minimal standard model, strictly interpreted, it is consistent to vary the strength with
which other fields couple to the $U(1)$ factor continuously.  That is to say, charge
quantization is not an automatic consequence of the gauge symmetry of the standard model. 
This flaw is mitigated by demanding anomaly cancellation, and removed in unified gauge
theories.)    Of course, these three parameters must describe {\it many\/} more
than three independent kinds of measurements.   Their success in doing so,
especially after the rigorous quantitative work of the LEP era, is truly remarkable.  
It is displayed in Figure~\ref{FWf2}.   As always, there are marginal misfits at the
edge of experimental and theoretical uncertainties, but no really convincing
discrepancy has emerged, despite much effort \cite{gambino}.  This impressive fit places very
severe constraints on ideas that ascribe electroweak symmetry breaking to the
influence of a new strongly interacting sector (technicolor), and exerts considerable
pressure on models where extra dimensions (``TeV gravity'', ``brane worlds'') \cite{csaki1} or complicated Higgs
sectors (``little Higgs'') \cite{csaki2} are invoked in this regard.

In his {\it Autobiographical Notes\/} Einstein describes the two sides of his
gravitational field equation $R_{\mu\nu} - \frac{1}{2} g_{\mu\nu} R = T_{\mu\nu}$
as constructions of gold and of wood.    His point, of course, is that the description of
gravity in terms of space-time curvature is profound and conceptually based,
whereas in his day the description of the energy-momentum of matter was by
comparison ramshackle and essentially phenomenological.  Were he with us today,
I think Einstein might be willing to say that QCD and the gauge sector of
electroweak theory are built from noble metals.

Not so the Higgs sector.     Within the standard electroweak model, in its minimal
form, it is the interactions of quarks and leptons with the Higgs field -- to be
precise, the interactions of matter with the Higgs condensate -- that are
responsible for their masses and weak mixing angles, including CP violation.   But
there is no deep principle in play.  The theory allows -- and the phenomenology
requires! -- a large number of independent parameters to describe these masses
and mixings.   It is a ramshackle structure, and we should certainly aspire to raze it
to the ground and replace it with something much better.

Though ramshackle, the standard model description of masses and mixings is
proving to be amazingly fruitful and durable.  Using it, Kobayashi and Maskawa
anticipated the existence of a third generation, so as to provide a mechanism to
accommodate the observed phenomenon of CP violation.  The third generation
promptly appeared.   Major experimental tests reported at this Conference, and
other recent work on CP violation and weak mixing angles, have failed to bring it
down.    I'll discuss this further below.  

\section{Completing the Standard Model: \\ The Higgs Particle}
\begin{figure}[htb]
\negsp
\centerline{\BoxedEPSF{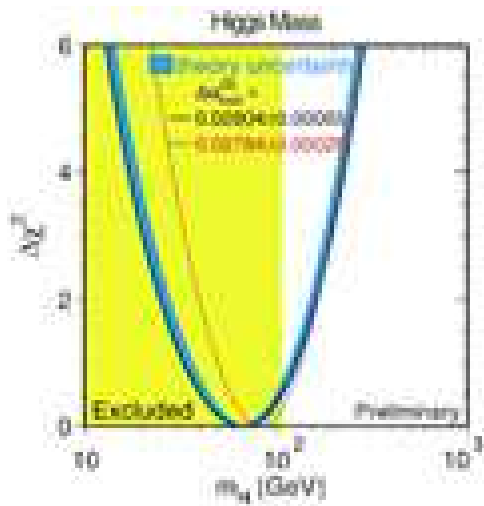 scaled 3000}}
\negsp
\caption{Estimates of the Higgs particle mass, assuming the minimal standard model, based on
precision electroweak measurements.   The preferred values lie near or below the
lower bound set by direct exclusion limits from LEP.   Values $m _H \gsim 150$
GeV are difficult to accommodate in this framework.}\label{FWf3}
\negsp
\end{figure}

\noindent The ultimate support for the ramshackle part of the standard model --
the wooden girders, so to speak, around which it hangs -- is the Higgs field. Its
independent physical quantum, the Higgs particle itself today remains the only
major ingredient of the Standard Model that has eluded  direct observation.  (To be
sure, the longitudinal parts of the $W$ and $Z$ bosons devolve from components of
the Higgs doublet!)  From the study of radiative corrections to electroweak
parameters, as indicated in Figure~\ref{FWf2}, one can infer limits on the Higgs
particle mass.  These limits assume, of course, that no additional unknown particles
are contributing significantly.   The limits are displayed in a more expansive format
in Figure~\ref{FWf3}.

\begin{figure}[htb]
\centerline{\BoxedEPSF{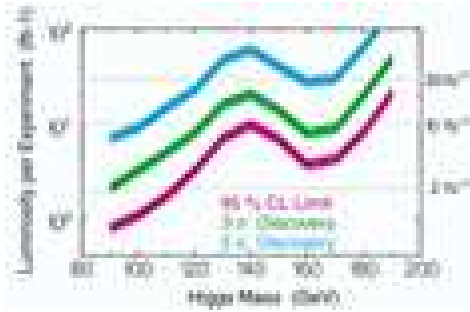 scaled 3000}}
\negsp
\caption{Estimates of the sensitivity of searches for the Higgs particles at the Tevatron, as a
function of $m_H$ and integrated luminosity.  Given luminosities of a few 10s of
fb.$^{-1}$, the range suggested in Figure 3 could be well covered.}\label{FWf4}
\negsp
\end{figure}

It is quite impressive and significant how well the mass is boxed in.   This makes the
challenge facing the Fermilab Tevatron, in searching for this particle, very tangible
and concrete.   That challenge is made graphic in Figure~\ref{FWf4}.

I'd like to take this opportunity to complain about an often repeated, but
nonetheless quite misleading, bit of hype to the effect that finding the Higgs
particle will explain The Origin of Mass.    First of all, the honest story of the true
origin of most of the mass of matter is extremely beautiful, and we ought to be
very proud of it and make it better known.   Most of the mass in ordinary matter
comes from its constituent protons and neutrons.  These in turn are built from 
nearly massless quarks and strictly massless gluons.   The mass of protons and
neutrons arises from the pure energy of QCD dynamics, according to
$m=E/c^2$.    We have calculated it, with quite respectable accuracy, based on a
theory that contains no continuous parameters whatsoever.  To my mind this ranks
among the very greatest achievements in all of science.    Second, as mentioned
above, the Higgs mechanism does not provide an even remotely satisfactory account of
masses and mixings; rather, these are accommodated using many continuous
parameters unconstrained by theory.  What the Higgs mechanism really explains
is not the origin of mass, but the breaking of electroweak symmetry.  Through it
we learn that empty space is a sort of exotic superconductor.   This accurate
representation of its nature is more challenging to explain, but ultimately more
profound and beautiful (not to mention true) than the usual vulgarization.  

With the opening of the LHC, the Higgs particle will become a much
easier target.   If low-energy supersymmetry is valid, there will be
several scalar ``Higgs particles'', including at least one charged and
two additional neutral species.   Sorting all this out will be a rich and
exciting enterprise, as suggested by Figure 5 \cite{bagger}.

In addition to the intrinsic importance and interest of its target,
the search for the Higgs particle involves some pretty points in
quantum field theory dynamics.   Uniquely, its primary coupling to
ordinary matter is entirely nonclassical; it communicates through a
virtual top loop to gluons!  Its primary production mechanism in
hadronic collisions is through gluon fusion.  This brings special interest
to the determination of gluon distribution functions, since they figure
directly into anticipating the production rate, and interpreting it once
measured.   Also, since the Higgs particle has vacuum quantum
numbers it can be produced with rapidity gaps \cite{ingelman}.  This will open
a new chapter in the ongoing saga of diffractive scattering, which has
been especially refreshed through recent discoveries at HERA.

\begin{figure}[htb]
\centerline{\BoxedEPSF{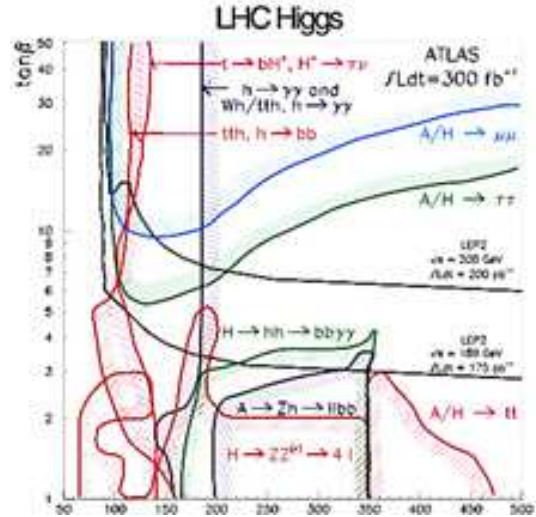 scaled 1800}}
\negsp
\caption{With the energies and luminosities anticipated at LHC, the standard
model Higgs is well covered, as is much of the range suggested for the additional Higgs
particles required to implement low-energy supersymmetry.}\label{FWf5}
\negsp
\end{figure}

\section{Unification and Supersymmetry}

\subsection{(Only Slightly In-)Direct Evidence}

\noindent
Because the standard model is so successful, and provides a close approximation to
the last word on Nature's inner workings,  we are obliged take its shortcomings
very seriously.   Consider Figure 6, which displays the bare-bones core of the
standard model, specifically the transformation properties of the lightest quarks
and leptons under the gauge groups $SU(3)\times SU(2)\times U(1)$.  
Left-handed fields are used exclusively, so we employ charge conjugation $u^c$ to
get the right-handed $u$ quark into the game, through its (left-handed)
conjugate.    $SU(3)$ acts horizontally, $SU(2)$ acts vertically, and the
hypercharge $U(1)$ assignments are indicated by subscripts.  Two shortcomings
hit the eye.  First, the particles fall into five disconnected pieces.  Second, there is
no evident rhyme or reason to the hypercharge assignments.  They are simply
chosen to accommodate experiment. Along the same lines, the gauge symmetry
falls apart into three independent pieces.
\begin{figure}[htb]
\centerline{\BoxedEPSF{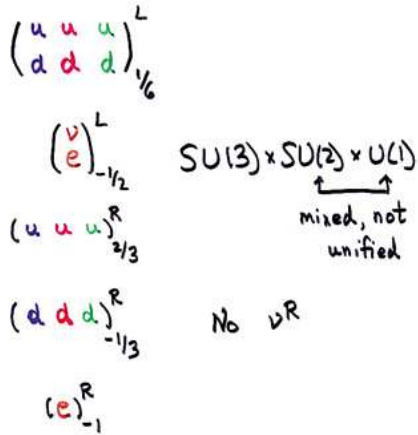 scaled 1000}}
\negsp
\caption{Multiplets in a single family of fermions the standard model.  Color
$SU(3)$ runs horizontally, weak $SU(2)$ runs vertically, and hypercharges are
indicated as subscripts.}\label{FWf6}
\negsp 
\end{figure}

These shortcomings can be overcome, in a way I find  compelling, by building upon
the concepts of the standard model itself.

\begin{figure}[hbt]
\centerline{\BoxedEPSF{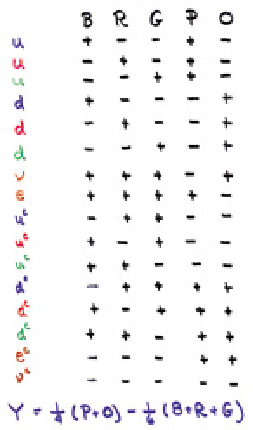 scaled 1300}}
\negsp
\caption{A single family of fermions fits neatly into an irreducible spinor {\bf 16}
of the unification symmetry $SO(10)$, a special and beautiful structure.   Besides the familiar
fermions, an additional singlet under
$SU(3)\times SU(2) \times U(1)$ is required.  It can be interpreted as a
right-handed neutrino, and plays an important role in the theory of neutrino
masses.}\label{FWf7}
\negsp 
\end{figure}

\subsubsection{Unification of Multiplets}
\noindent
The gauge symmetry $SU(3)\times SU(2)\times U(1)$ and the observed
fermions fit snugly into an $SU(5)$ unification.  Using a simple breaking scheme (condensate in
the adjoint {\bf 24} representation), and starting with fermions in the antisymmetric
tensor $\overline{\bf 10}$ and vector {\bf 5} representations, we arrive at
precisely the gauge groups and fermion multiplets of the standard model,
including the hypercharge assignments. This is a highly nontrivial coincidence. 
Since it cuts the number of multiplets down from five to two, and uniquely fixes
the hypercharge assignments, this unification achieves substantial esthetic gains
over its starting point. Still more beautiful is the possibility of unification afforded
by the slightly larger group $SO(10)$.  Now the fermions all fit into a single spinor
{\bf 16} representation.  This is a particularly elegant representation, with
remarkable properties, as indicated in Figure~\ref{FWf7}.  The components of the
spinor representation can be specified by their transformation properties under
the diagonal $SO(2)\times SO(2)\times SO(2)\times SO(2)\times SO(2)$.  These have
the physical interpretation of being the  values of five color charges.   All possible
combinations of charges $\pm \frac12$ are allowed, subject to the constraint
that the number of $+\frac12$ charges is even.  From these abstract
mathematical rules, the multiplet structure of a complete family in the standard
model falls out automatically, matching the pattern observed in Nature.   In
particular, the hypercharges are uniquely predicted from the strong and weak
charges, according to the simple formula
$$ 
Y~=~ -\fract16 (B+R+G) + \fract14(P+O)~.
$$

The spinor {\bf 16} contains, in addition to the fermions of the standard model,  an
additional  particle $N$.   $N$ is a singlet under $SU(3)\times SU(2)\times U(1)$,
and so it has none of the standard gauge interactions with matter.  Its
``non-observation'' does not pose immediate problems.  Indeed it plays a major {\it
constructive \/}  role in the theory of neutrino masses, as I shall discuss presently.

\subsubsection{Unification of Couplings}
\noindent Unified gauge symmetry requires universal gauge coupling strength. 
This does not hold, of course, in the standard model.    The $SU(3)$ coupling is
observed to be larger than the $SU(2)$ coupling, which in turn is larger than the
$U(1)$ coupling.

Fortunately, as we have seen in Figure~\ref{FWf1}, a great lesson from QCD is that
coupling constants evolve with energy.  The same sorts of calculations that give us
asymptotic freedom in the strong interaction allow us to evolve, theoretically, the
effective couplings up to high energy, or equivalently short distance, scales.   If
the gauge part of the standard model derives from a larger gauge symmetry,
spontaneously broken at a unique large energy scale, we should expect that these
couplings meet at a point.  For in their evolution from high to low energies, the
couplings only started to diverge after the big symmetry was broken.

\begin{figure}[hbt]
\centerline{\BoxedEPSF{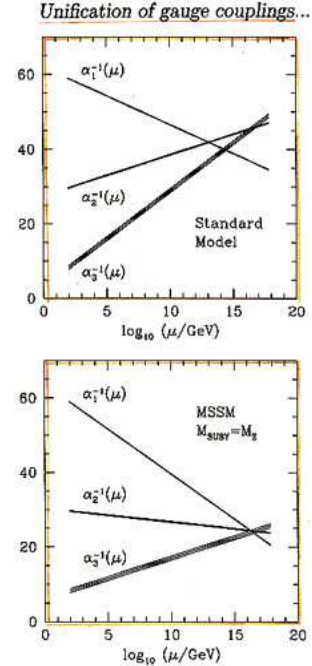 scaled 1000}}
\negsp
\caption{Calculation of the running of couplings, based on the renormalization
group.   To a first approximation, the inverse couplings run logarithmically with
the energy scale.  The hypercharge is normalized as required by unified theories based on
$SU(5)$, $SO(10)$, or their most straightforward variants.  In the top panel the
effects of virtual particles in the minimal standard model are included; in the
bottom panel the effects of virtual particles in its minimal extension to incorporate
low-energy supersymmetry are included.  The former doesn't quite work; but the latter
appears quite viable.}\label{FWf8}
\negsp\negsp
\end{figure}

If we evolve the couplings up to high energy using only the particles of the
minimal standard model, we get the result shown in top part of Figure~\ref{FWf8}. 
Notice that to a good approximation the inverse couplings are predicted to run
logarithmically, so the running generates straight lines in this log plot. The width of
the lines indicates the experimental uncertainties, post LEP.  It is a remarkable
near-miss, but a miss nonetheless. One can try to repair this small discrepancy in
any number of ways, using various slight perturbations on the minimal model.   
In the absence of any powerful guiding principle, however, such fixes lack
conviction.

Instead of tinkering with the standard model, let us consider the apparently drastic, but
independently motivated, idea that supersymmetry is broken only at relatively
low ($\lesssim$ TeV) energies.   This modifies the running of the couplings, because
there are more virtual particles to consider,  in a way that is easy to compute. 
Though it involves a vast extension of the minimal standard model (more than
twice the particles!), low-energy supersymmetry has a surprisingly small, and
remarkably salutary, effect on the unification of couplings calculation.   If we
extend the standard model in the most economical way to include low-energy
supersymmetry, we find the result shown in the bottom part of
Figure~\ref{FWf8}.    The unification now works much better, quantitatively.    This
is an extremely encouraging result, both for unification and for low-energy
supersymmetry.  

The quantitative success of the unification of couplings calculation (with
low-energy supersymmetry) is undeniable.  How seriously should we take it?

At the most formal level, the unification of couplings is an over-constrained fit of
three measured quantities -- $\alpha_1(M_{\mathrm W})$, $\alpha_2(M_{\mathrm
W})$, $\alpha_3(M_{\mathrm W})$ -- to two theoretical parameters, the scale of
unification and the strength of coupling at unification.  Given the precision of the
measurements, it is remarkable that an adequate fit is obtained.

But simply saying that one number falls into place does not do justice to the state
of affairs, because many other things, besides failure to
satisfy this one numerical constraint, could have gone wrong.  If the couplings had met at too
small an energy scale, we would have difficulties with rapid proton decay.   If they had met
at a significantly larger a mass scale, at or above the Planck scale, we would have
had to worry about quantum gravity corrections.   The actual scale at which they
meet, not far on a logarithmic scale, but still significantly below, the Planck scale, is
uniquely acceptable.  Similarly, if the unified coupling were much larger we could
not trust the perturbative calculation.

To me, the unification of multiplets and the unification of couplings are the crown
jewels in an inventory of physics beyond the standard model. Together, they make
a powerful {\it prima facie \/} case for the elements that went into their derivation:
unified gauge symmetry, for the unification of multiplets; renormalizable quantum
field theory, operating smoothly up to near-Planckian scales, for the proper
logarithmic running of couplings; and low-energy supersymmetry, for detailed
numerical success. Nowadays, in the context of string theory, we know -- or,
rather, we have incomplete suggestions about -- many alternative ways that the
low-energy
$SU(3)\times SU(2)\times U(1)$ symmetry of the standard model might emerge
from constructions that involve  neither effective unified gauge field theories nor
symmetry breaking through condensates.    Of course there is no necessary
contradiction, since early reduction to an effective unified gauge field theory also
still remains a viable option.   Along this line, perhaps we should take the striking
success of the ``oldies but goodies''  I've just recalled as indications that
in searching for string-based models of Nature, we should look with favor upon
those that reduce to something like an effective supersymmetric $SO(10)$
renormalizable gauge field theory, or a recognizably broken version thereof, just
below the Planck scale.   At least we can say that other schemes have some
coincidences to explain.

\subsection{Additional Evidence, \\ For the Sympathetic}

The unification of couplings calculation is unique in its 
{\it quantitative\/} success, but in addition there are several
impressive {\it qualitative\/} points in favor of low-energy
supersymmetry.   

The most profound concerns stabilization of the electroweak scale.  In
the minimal standard model, and in a generic extension of it, radiative
corrections to the Higgs field mass parameter, which governs the scale
of electroweak symmetry breaking, are quadratically divergent.   In
order that these corrections not dwarf the final value we need, the
cut-off must be imposed at
$\lsim 1$~TeV.   But we would like to contemplate physics at much
higher scales, and did so with striking success in the unification of
couplings calculation, so it is preferable to have these corrections
cancel.  By balancing off virtual bosons and fermions, supersymmetry
accomplishes this cancellation.  Supersymmetry effectively broken at
$\lsim 1$ TeV will protect the electroweak scale adequately.

Other mechanisms have been proposed to generate or stabilize the
electroweak symmetry breaking scale.  But, as I mentioned previously,
they tend to leave nontrivial footprints in the form of radiative
corrections, and there is little sign of
deviations from the standard model in precision electroweak
measurements.   Another important advantage of low-energy
supersymmetry is that the radiative corrections it generates are
generally small.   Figure 9 displays this feature.

\begin{figure}[hbt]
\negsp
\centerline{\BoxedEPSF{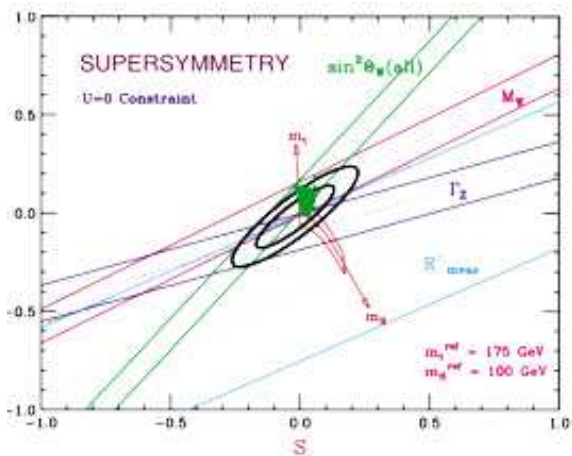 scaled 1600}}
\negsp
\caption{Modification of the precision electroweak parameters in models
incorporating low-energy supersymmetry.  Because these models contain several 
parameters that relevant, but currently undetermined, a sampling of results from typical models
in the allowed space is displayed, as the solid blob.  The important point is that the
corrections to the standard model are generically small, and lie comfortably within existing
experimental limits.}\label{FWf9}
\negsp 
\end{figure}

Furthermore, standard model fits to precision measurements
require a low value for the Higgs mass, near (or below!) existing
experimental limits.   In the standard model itself, there is no reason
to favor such a value.   But in its extension to incorporate low-energy
supersymmetry, the Higgs mass is tied to the $W$ and $Z$ masses,
and it must be light.

\subsection{Where It Takes Us}

\noindent

Unless all these indications are part of an elaborate, cruel jest on the
part of the Creator, we can look forward to a golden age of
discovery.   New quantum dimensions of superspace will
open up, inhabited by weird {\it doppelgangers\/} of the familiar fundamental particles.

Both the strength and the weakness of the unification of couplings calculation is its robustness.  Because the inverse couplings run logarithmically,
factor-of-few reshufflings in the mass spectrum of the contributing particles tend to induce only small changes in this calculation.  That feature, of course, is
what allows us to abstract a general success for low-energy supersymmetry and unification, which is fairly
insensitive both to the actual spectrum of supersymmetric particles and to the details of unified symmetry
breaking.   The other side of the coin is the implication that this success does not provide much resolving
power for those details.  

The mechanism of supersymmetry breaking is still up for grabs, with several proposals under active consideration.  They lead to mass spectra with quite
distinctive patterns, as displayed in Figure 10.   These mechanisms could encode, respectively, the first tangible influence of quantum gravity in subatomic
physics (gravity mediation) the existence of new strongly interacting sectors (gauge mediation), dynamics from small curled-up extra spatial dimensions
(anomaly and gaugino mediation), or combinations of these.  Exciting stuff!

\begin{figure}[hbt]
\centerline{\BoxedEPSF{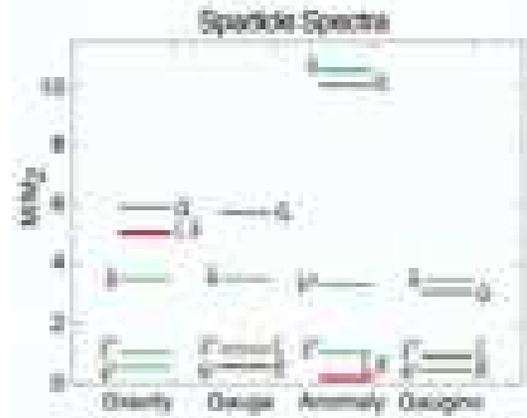 scaled 3600}}
\negsp
\caption{Cartoon depiction of characteristic mass patterns that are obtained in
different proposals for supersymmetry breaking.   The important message is that
measurement of the spectrum will allow us to distinguish among these
mechanisms.}\label{FWf10}
\negsp 
\end{figure}

Low-energy supersymmetry provides excellent candidates for the dark matter that cosmology seems to demand. 
The density of dark matter produced depends on poorly conditioned details of the model but, as displayed in
Figure 11, there is a healthy swath of parameter space where it roughly matches what the cosmologists want.  
Also indicated in this Figure are some of the many experimental approaches to identifying dark matter of this
kind.

\begin{figure}[hbt]
\centerline{\BoxedEPSF{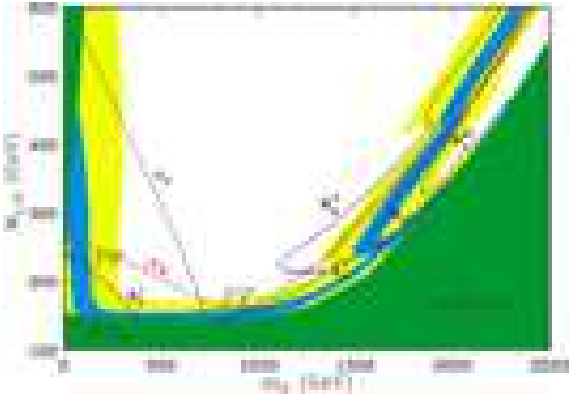 scaled 3200}}
\negsp
\caption{Dependence of supersymmetric dark matter production on two of the
standard parameters for low-energy supersymmetry breaking.  In the light blue
(or light gray!) band, the lightest $R$-odd particle supplies all or most of the
density required by astronomical observations.  Also indicated are expected
sensitivities of various experimental probes.  For a full explanation, see \cite{feng}.}\label{FWf11}
\negsp 
\end{figure}

Finally, I should own up to the dark side of low-energy supersymmetry.   It offers
many new potential sources of flavor and CP violation, including baryon number violation.  Some of these are associated with low-dimension
operators, so that {\it a priori\/} they are sensitive to physics at high mass scales, where exotic effects of quantum gravity and exchange of the new
particles associated with unification are unsuppressed.  Nature does not seem to avail Herself of these possibilities, at least not that we've seen so
far.   So special mechanisms and symmetries must be postulated, to keep the basic ideas of low-energy supersymmetry and unification
phenomenologically viable.   Possibilities have been suggested, so it's not an outright contradiction (for example, gauge mediation cleanly suppresses
the potential flavor and CP  violation), but we need more complete and convincing ideas.  Interpreting things optimistically, this is a relatively
poorly explored area which makes close contact with some of the most fundamental aspects of supersymmetry, unification, and
string theory, so  major new theoretical insights might still be plausibly expected.    And several concrete estimates suggest that
positive experimental discoveries lie just ahead in electric dipole moments, $\mu-e$ conversion processes, and proton decay.

\section{Segue}

\noindent
Now  I need to shift gears, and discuss recent progress in a few areas where
experiments are supplying us with wonderful results, but results that we can't 
yet do justice to theoretically.    At this point, a joke is in order.

A man walks into a bar, takes a seat on the next-to-last stool, and spends the
evening chatting up the empty stool next to him, being charming and flirtatious, as
if there were a beautiful woman in that empty seat.  The next night, same story. 
And the next night, same story again.  Finally the bartender can't take it any
more.  She asks, ``Why do you keep talking to that empty stool as if there were a
beautiful woman in it?'' 

The man answers, ``I'm a theoretical physicist.  I'm hoping that a beautiful woman
will tunnel in from an extra dimension and materialize on that stool.   Then I'll
seem very clever indeed, and  I'll have the inside track with her.'' 

``That's ridiculous,'' says the bartender.  ``Plenty of  very attractive women come
to this bar all the time.  You're reasonably presentable, and extremely articulate; if
you applied your charm on one of them, she might be interested in you.''

``Oh come on,'' he replies, ``how likely is {\it that\/}?''

\section{Results in Flavor Physics}

\subsection{CP Violation}

This is a vast and intricate subject, with a lot of relevant data gushing in right now  from BABAR and BELLE on $B$ meson physics, as well as final
results from heroic, decades-long programs at CERN and Fermilab measuring $\epsilon^\prime/\epsilon$.  
Fortunately for me the results and their interpretation were beautifully reviewed at this Conference by Yosi Nir
\cite{nir}, and I don't have much to add.   Perhaps the most eloquent thing for me to do is simply to display Figure
12.   The delicate relations of the unitarity triangle, which are overconstrained by the data, appear to be well
obeyed.   The dominant source of CP violation in the $B$ and $K$ systems thus appears, on the face of it, to derive
from an irremovable complex phase appearing in the mixing matrix for 3 quarks, just as Kobayashi and Maskawa
proposed.   The intrinsic phase is not small; the relative smallness of CP violation in the $K$ meson system, which for
many years is all we've had to look at, is because this system is fairly well insulated from goings on in the heavy
quark sectors.  

Of course, more complete and accurate measurements may still reveal subtle deviations from this picture, but for
now the ramshackle flavor structure of the minimal standard model is holding up.

\begin{figure}[hbt]
\centerline{\BoxedEPSF{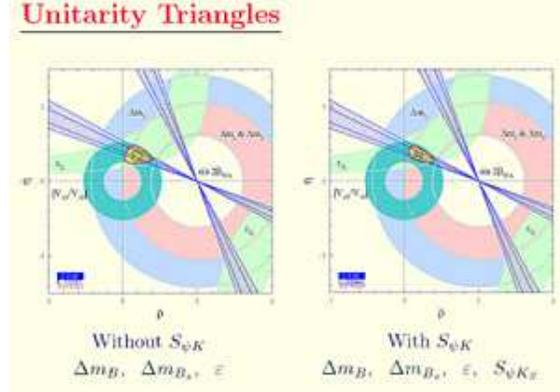 scaled 1400}}
\negsp
\caption{Unitarity triangles, derived from orthogonality relations in the
3-family Cabibbo-Kobayashi-Maskawa unitary mixing matrix.  If this framework is
adequate to describe the processes measured, the various allowed bands must
overlap, thus marking out a consistent, allowed region for the underlying
parameters.   So far, they do.  For more details, see \cite{nir}}\label{FWf12}
\negsp 
\end{figure}

Before leaving the subject, I'd like to express my admiration for the beauty of this physics.   In Figures 13 and 14, lifted from Nir, you see how the
relative phases between very different sorts of amplitudes interfere to govern various kinds of physical processes, and the intricacy of the resulting
formulas.  This complex of ideas and measurements could be used as the basis of a course in fundamental quantum mechanics -- and provides an
extremely impressive demonstration, in an extremely exotic setting, of how well it works!   Through familiarity we can easily lapse into taking that for
granted, but it is an extraordinary fact.  

\begin{figure}[hbt]
\centerline{\BoxedEPSF{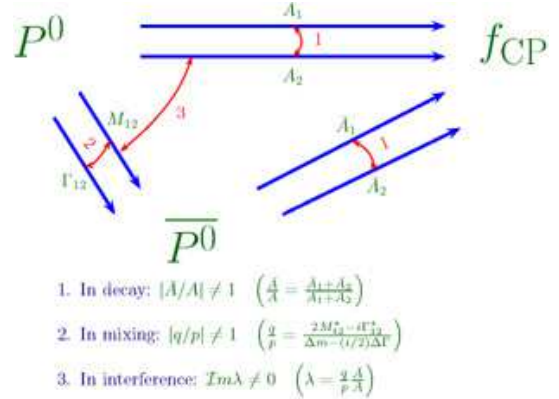 scaled 1300}}
\negsp
\caption{Indication of how different sorts of amplitudes contribute coherently to
the value of observable parameters in different sorts of CP violation experiments.}\label{FWf13}
\negsp  
\end{figure}

\begin{figure}[hbt]
\centerline{\BoxedEPSF{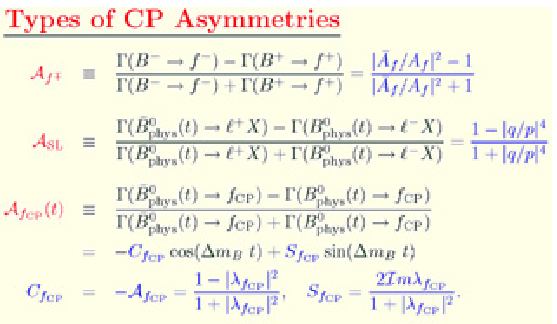 scaled 1400}}
\negsp
\caption{Some of the intricate formulas that result from the application of
quantum mechanics in the situations described by Figure 13.}\label{FWf14}
\negsp 
\end{figure}

\subsection{Neutrino Oscillations}

This is another vast and intricate subject on which there has been dramatic progress recently.  

We've just heard a nice summary of the experimental situation, and there are many reviews available, so again my main duty is simply to recall the
appropriate image, shown as Figure 15 \cite{gonzalez}.   The existence of neutrino oscillations, with large mixing angles, is
secure.  Neutral current results from the SNO collaboration, coming in slightly after the Conference, confirmed that the Sun is
putting out its full share of neutrinos, and the long-standing deficit of electron neutrinos observed in charged currents is
mostly, and presumably entirely, due to oscillations.    The various solar neutrino experiments all appear to be fit very well
to the so-called large mixing angle, or LMA, solution.

\begin{figure}[hbt]
\centerline{\BoxedEPSF{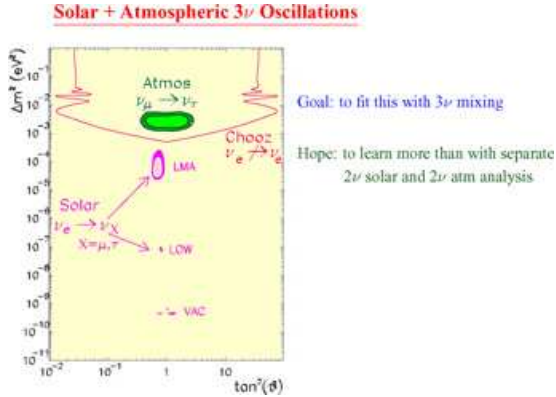 scaled 1200}}
\negsp
\caption{Summary of results on neutrino oscillations.  All the experiments, with one exception
(LSND) are consistent with a simple picture based on masses and mixings among the three
known neutrinos.  From atmospheric neutrino oscillations we infer a mass$^2$ difference of
order
$5\times 10^{-3}$ eV$^2$ and near-maximal mixing; from solar neutrino oscillations we
can apparently infer the so-called large-mixing-angle (LMA) solution, as indicated.}
\label{FWf15}
\negsp\negsp
\end{figure}

Within the framework of
electroweak $SU(2)\times U(1)$, neutrino masses can be accommodated by means of a
nonrenormalizable interaction
$$
\Delta {\cal L} ~=~ {1\over M} \phi^\dagger l  \phi^\dagger l 
$$ where $l$ is the lepton doublet and $\phi$ the Higgs doublet.  With $\phi$
replaced by its vacuum expectation value $v$, this becomes a Majorana
neutrino mass of magnitude
$v^2/M$.   With $M$ of order $10^{15}$--$10^{16}$ GeV, this is about right.  That
mass scale is equal to what appears in the unification of couplings
as the
scale at which unification symmetry breaks.

There is a simple, concrete dynamical mechanism for generating neutrino
masses which explains this coincidence.   It involves the $N$ particle we met
before as the missing component of the $SO(10)$ spinor {\bf 16}.  Since it is
an $SU(3)\times SU(2) \times U(1)$ singlet, this particle can acquire a large mass
$\sim M_{\rm U}$ at the scale where $SO(10)$
symmetry is broken, without breaking those low-energy symmetries.  It also can connect to the conventional left-handed
neutrino $\nu$ by a normal Higgs-type mass term acquiring a mass $m$.  By
second-order perturbation theory, passing through the intermediate $N$, we
generate a Majorana mass of order $m^2/M_{\rm U}$ for $\nu$.   Finally, we
might expect $m\sim v$ for the heaviest neutrino, if this mass is related by
symmetry to the large top quark mass.

There are significant uncertainties at both stages in calculation, so this
accommodation of the scale of neutrino masses is at best semiquantitative.
Still, I find it  very impressive how the outlandishly small value of the neutrino mass,
relative to other quark and charged lepton masses, gets mapped to the
outlandishly small value of the unification scale, and how the existence of
$N$, at first sight an embarrassment, turns out to be a blessing.

\subsection{What Does It All Mean?}

\noindent
So what does it all mean?  That's a rather embarrassing question, I'm afraid.  

We've learned a few important lessons.  The ramshackle part of the
standard model holds up amazingly well.  CP violation in the quark sector is not intrinsically small.  There is a pronounced mass hierarchy for quarks,
with small mixing, and (probably) a significant mass hierarchy for neutrinos, with large mixing.   The magnitude of neutrino masses roughly
accords with expectations from unified gauge theories.

The biggest issues, however, remain the obvious ones, and they remain
unresolved.   Why is there repetition of families? Why are there three?  Why is
there such a spread in quark and lepton masses?   There is a factor of roughly
$10^{-6}$ between the electron mass and the top quark mass, a tiny number that begs for a qualitative explanation, but we have
only vague speculations about its origin.  Can we infer profound symmetries, or
profound dynamical principles, from the study of flavor?  Or will it remain,
indefinitely, a spectacularly abstruse branch of natural history?

Some familiar facts seem especially clear and significant.  Masses, like couplings, evolve with energy scale, and
the observed ratio $m_b/m_\tau$ is at least roughly consistent with equality at the unification scale.  Although
for historical reasons we usually speak of the $t$ quark as being extraordinarily
heavy, there are good reasons to consider $m_t$ as the most reasonable of quark
masses.    Indeed, a wide range of masses at the unification scale focus down to
roughly the observed value of $m_t$ at accessible energies.   

A sophisticated and possibly important effort in model-building is directed toward
forging links between the pattern of masses and mixings and the one hand and
gauge unification on the other \cite{barr}.    The light Higgs doublet (or doublets), whose
vacuum expectation value generates the masses  and mixing, can descend from a
combination of irreducible representations of the unified symmetry.  Those
representations have different restrictions on interfamily couplings, e.g.,
symmetry or antisymmetry in the family indices; and within a family, different
ratios for their contributions to quark versus leptons mass matrices.  By using {\it
all\/} the data, and looking for patterns, we can start to construct a genealogy.  

This quest ties up with the question of why the light Higgs doublet of electroweak theory is badly split  from
its unified partners \cite{dine}.  Those partners must be extremely heavy, as otherwise they would 
mediate unacceptably large contributions to proton decay.  There are fairly simple mechanisms that  explain, or at least
naturally accommodate,  this splitting; as one might expect from the heuristic, bottom-up argument about stabilizing the
electroweak scale, one thing these mechanisms have in common is that they rely heavily on special properties of
supersymmetric theories.

Since in these schemes quark and
lepton mixings are intimately related, one might suspect that
tension arises between the smallness of quark mixings and the largeness of lepton
mixings revealed in neutrino oscillations.    There is no conflict, but one is led along
these lines in surprisingly specific directions, involving so-called ``lopsided'' mass
matrices \cite{barr}.    A generic prediction of this framework is that the mixing observed in
atmospheric neutrino oscillations will not be accurately maximal.    

A seemingly very different approach attempts to forge links between the geometry of small folded up extra dimensions and the observed pattern
of masses and mixings.   Some ideas here are that fermions in different families live at different places in the extra dimensions (e.g., that they are
localized on orbifold singularities), that the Higgs field has different amplitudes at these places, which accounts for their different masses, and that
overlap of their wavefunctions determines the mixings.   
Some attractive, reasonably economical models have been constructed along these lines \cite{jmr}.  

Of course, even successful model-building and pattern-matching of this kind will
not satisfy our hunger for deep insight into the physical world.  We want to know
what big ideas are in play, {\it why\/} things are the way they are.    What principles determine which unified symmetry multiplets condense, and
which combinations stay light to become the electroweak Higgs doublets?  What dynamics determines the shape of the extra
dimensions, and where the fermions live?  Is there a useful notion of symmetry among the families, and if so how is it broken,
and how could we tell?  By answering questions of this order we might bring the subject to an appropriate, higher level.

To sum up: despite much hard and extremely impressive work, and some
significant progress, our understanding of fermion masses and mixings remains
superficial at best; we are still at the level of pattern-gathering, similar to where we were in strong
and weak interaction physics in the early 1960s (or maybe the
1910s?).   Fortunately, there are real prospects for gathering important new
information, and observing fundamentally new phenomena, soon.   Many avenues are ripe for exploration: electric dipole
moments, charged lepton number violation, baryon number violation, and rare decays in $K$ and in
$B$ physics.   On the theoretical side, we desperately need more powerful, sharply
testable ideas.  

\section{Cosmic Connections}

\subsection{Challenges from Cosmology}

Measurements of cosmological parameters have improved dramatically in recent years, primarily due to measurements of cosmic microwave
background anisotropies and supernova redshift surveys.   The supernova part of this is summarized in Figure 16, taken from Marc
Kamionkowski's excellent review here.  

\begin{figure}[hbt]
\centerline{\BoxedEPSF{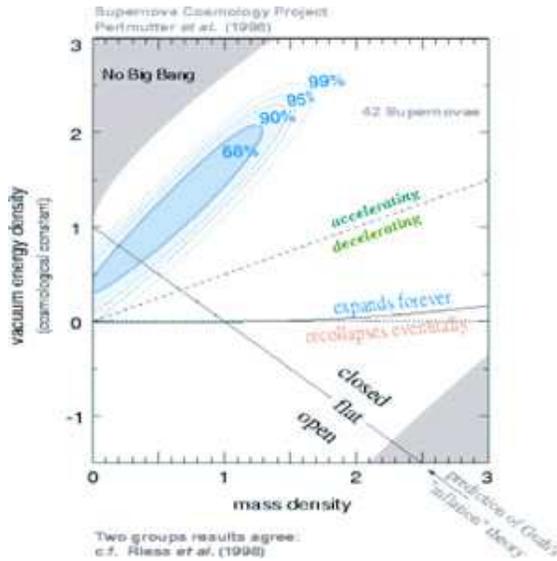 scaled 1400}}
\negsp
\caption{Indications of the mass density of the Universe, and the amount of
negative pressure (``vacuum energy density'') based mainly on supernova
surveys.   Together with these, measurements of microwave background
anisotropies and direct mass surveys, converge on a simple but surprising
inventory of mass in the Universe, as discussed in the text.}
\label{FWf16}
\negsp \negsp
\end{figure}

All the measurements seem to converge upon a simple, but unexpected and challenging picture.  The answer is sufficiently weird that we might be led
to rethink the foundations, but for now Einstein's general relativity and the Friedmann-Robertson-Walker model of a homogeneous
and isotropic expanding universe, with today's structure emerging from the growth of small perturbations early by gravitational
instability, form at least an adequate descriptive framework, which I'll adopt.  The universe is observed to be very nearly
spatially flat, and accelerating.   Spatial flatness is correlated with a critical value of the average mass density, $\rho_c = 
3H^2/8\pi G$, with $H$ the Hubble expansion parameter.  In units where the critical density is 1, ordinary (baryonic) matter
contributes about $\rho_B \sim 0.03$, some unknown pressureless `dark matter' contributes $\rho_d
\sim 0.3$, and some negative-pressure `dark energy' contributes $\rho_d \sim 0.7$, with $p_d \approx - \rho_d$.  

Plausible candidates for the dark matter include a quasi-stable particle served up by low-energy supersymmetry -- the lightest particle with
negative
$R$-parity, where $R= (-1)^{B+L+2s}$ is 1 for all standard-model particles and $-1$ for their superpartners -- and the axion, to be
discussed below.   There are important, vigorous searches underway to pin down these possibilities.  

The properties of the dark energy are consistent with its deriving from a cosmological term.   Depending on whether we put this on the left-hand
side of Einstein's equation together with the curvature, or on the right-hand side together with the energy-momentum tensor, we can think of it
either as a modification of the basic equations of gravity or as a peculiar form of matter.   However regarded, its actual value is difficult to reconcile
with other aspects of our understanding of Nature.  

On the one hand, its value seems absurdly small.  Our best-established theories in the standard model require that empty space is permeated with
condensates -- at the very least a condensate for spontaneous chiral symmetry breaking in QCD, and another for electroweak symmetry breaking. 
Naive estimates of the contribution of these condensates to the energy yield values about 60 orders of magnitude larger than the observed value. 
And condensates associated with supersymmetry breaking should be, and unified symmetry breaking could be, even heavier.   But gravity, which is
universal, seems to blithely ignore all this structure, as well as energy that might be associated with zero-point motion of quantum fields.  

On the other hand, its actual non-zero value seems weirdly coincidental.   Why should the weight of empty space, which presumably ought to be
determined by local physics, have anything to do with the density of (dark) matter?  If it really is a cosmological term, this is a coincidence that will
not stand the test of time, since the matter density gets diluted by the expansion of the
Universe while the cosmological term abides; nor was it true in the very early Universe.    It is very suspicious to label as
coincidence a quantitative correlation that happens to be true in every case we actually measure it  -- even if, as here, the
number of such cases is one.   Maybe it indicates a flaw in the foundations.  

To me, this is the  biggest mystery in physical science.  It makes a mockery of
claims that we have a ``Theory of Everything'' or a ``consistent theory of quantum gravity''.  Everything -- but not including most
of the mass of the Universe? Consistent  -- but not with one of the most fundamental qualitative features of gravity in Nature?

Putting these deep concerns aside, the emerging picture is strikingly consistent with expectations from inflationary models.  A period of rapid
inflation in the early universe can very naturally explain its flatness.   Detailed models, coming in several varieties 
(slow-roll, chaotic, eternal, hybrid, etc.) also suggest a mechanism for generating the fluctuations that seed structure formation that
is consistent with information emerging from the measurement of microwave background anisotropies.  At present the tests
are quite weak, but they could be greatly strengthened by future measurements of polarization.   

Assuming that some simple inflationary model survives rigorous testing, we shall be faced with a big theoretical challenge.  Essentially all  current
models are based on postulating the existence and properties of ``inflation'' fields {\it ad hoc\/} -- and, of course, extrapolating
the cosmological term back in time naively.  On the face of it they all involve fine-tuning, invoking small couplings to explain the
drawn-out period of inflation and the smallness of fluctuations.  It would be very satisfying to identify appropriate fields based on
considerations of fundamental physics.   Alternatively, we might be required to find mechanisms that achieve the existing successes,
which are basically qualitative, in a different way.   

\subsection{Cosmic Rays}

High-energy cosmic rays are interesting for many reasons, of course, in astrophysics.   They extend to higher energies than will be
available at accelerators in the foreseeable future, so it is important to watch for lessons they might contain about fundamental
physics.   A new generation of powerful detectors, prominently including the Auger array, should soon bring the study of this domain
to a new level.  These matters were nicely reviewed by Tom Gaisser.  

For many years the most striking anomaly in the field has concerned the most energetic cosmic rays.  This is the existence of significant
numbers of events with primary energy above about $10^{11}$ GeV, the
so-called GZK cutoff.   Conventional particles --  protons, photons, and nuclei --
above this energy have a hard time propagating over cosmic distances, so one
expects the spectrum of cosmic ray primaries to decrease sharply just below this
energy.   It has appeared that this was not the case, that the spectrum held up or
even rose.  This has inspired several proposals for exotic particle-physics sources,
which I won't describe in detail here except to say that all have
serious problems.  

Recent recalibrations, displayed in Figure 17, now cast some doubt on the existence of the anomaly.  It is too soon to tell how this
situation will resolve itself -- Auger should give decisive experimental results -- but it is a bit of news worth noting.  

\begin{figure}[hbt]
\centerline{\BoxedEPSF{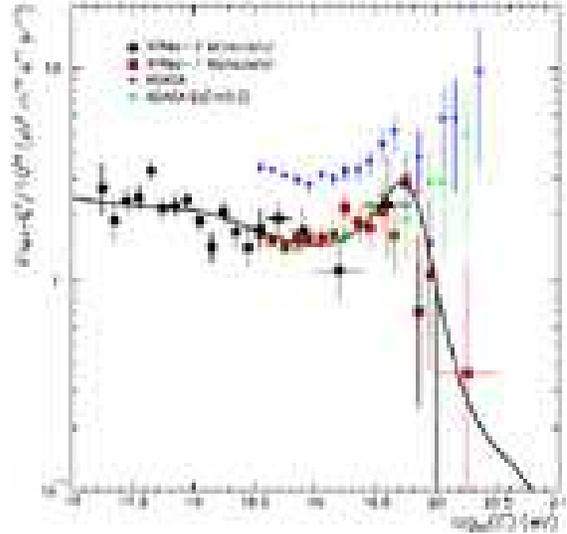 scaled 2900}}
\negsp
\caption{The latest calibration of HiRes data (round and square dots) is not inconsistent with a
conventional spectrum, including the GZK cutoff, as indicated.  Results from AGASA disagree
with these both below and above the cutoff.  An overall re-scaling of the energy estimates by
$\sim 20$\% could bring them into reasonable agreement.}
\label{FWf17}
\negsp 
\end{figure}

\section{An Ultralight Sector?}

Many ideas suggest the possible existence of a new sector of physics, consisting of very weakly interacting, ultra-light or even massless
spin-0 particles.   These have various names, including axions, familons, modulons, vadrons (a new one; see below), and dilatons.   
Aside from the dilaton, which is a special case, they are associated with spontaneous symmetry breaking, essentially as the
Nambu-Goldstone bosons of various possible broken symmetries.  The axion is the best motivated and best developed, so I'll focus on it
as representative.

\subsection{Recollections of Axions}

Given its extensive symmetry and the tight structure of relativistic quantum field
theory, the definition of QCD only requires, and only permits, a very restricted set of
parameters.  These consist of the coupling constant and the quark masses, which
we've already discussed, and one more -- the so-called
$\theta$~parameter.  Physical results depend periodically upon $\theta$, so that
effectively it can take values between $\pm \pi$.  We don't know the actual value of
the
$\theta$ parameter, but only a limit, $|\theta | \lsim 10^{-9}$.   Values outside this
small range are excluded by experimental results, principally the tight bound on the
electric dipole moment of the neutron.   The discrete symmetries P and T are
violated by  $\theta$ unless $\theta \equiv 0$ (mod $~\pi$).   Since there are P- and
T-violating interactions in the world, the $\theta$ parameter cannot be put to zero
by any strict symmetry assumption.   So its smallness is a challenge to understand.  

The effective value of $\theta$ will be affected by dynamics, and in particular by
condensations (spontaneous symmetry breaking).   Peccei and Quinn discovered
that if one imposed a certain asymptotic symmetry, and if that symmetry were
spontaneously broken, then an effective value
$\theta \approx 0$ would be obtained.  Weinberg and I explained that the approach
$\theta
\rightarrow 0$ could be understood as a relaxation process, wherein a very light
collective field, corresponding quite directly to $\theta$, settled down to its
minimum energy state.  This is the axion field, and its quanta are called axions.  

The phenomenology of axions is essentially controlled by one parameter, $F$.  $F$
has dimensions of mass.  It is the scale at which Peccei-Quinn symmetry breaks. 
More specifically, there is some scalar field $\phi$ that carries Peccei-Quinn charge
and acquires a vacuum expectation value of order $F$.  (If there are several
condensates, the one with the largest vacuum expectation value dominates.)  The potential
for $| \phi |$ can be complicated and might involve very high-scale physics, but the
essence of Peccei-Quinn symmetry is to posit that the classical Lagrangian is
independent of the phase of $\phi$, so that the only way in which that phase affects
the theory is to modulate the effective value of the
$\theta$ term, in the form $\theta_{\rm eff.} = \theta_{\rm bare} + ~{\rm
arg}~\phi$.\footnote{I am putting a standard integer-valued parameter, not discussed here, 
$N=1$, and slighting several other inessential
technicalities.}  Then we identify the axion field
$a$ according to
$\langle \phi \rangle \equiv  F e^{ia/F}e^{-i\theta_{\rm bare}}$, so $\theta_{\rm eff.} = 
a/F$.   This insures canonical normalization of the kinetic energy for
$a$.  

In a crude approximation, imagining weak coupling, the potential for $a$ arises
from instanton and anti-instanton contribution, and takes the form
${\frac12}(1-\cos \theta_{\rm eff.})\times\linebreak
e^{{-8\pi^2}/{g^2}}\Lambda_{\rm QCD}^4$.    So the
energy density controlled by the axion field is 
$e^{{-8\pi^2}/{g^2}} \Lambda_{\rm QCD}^4$.  The potential is minimized at
$\theta_{\rm eff.} =0$, which solves the problem we started with.    The mass$^2$
of the axion is
$e^{{-8\pi^2}/{g^2}} \Lambda_{\rm QCD}^4/F^2$.  Its interactions with matter
also scale with $\Lambda_{\rm QCD}/F$.  The failure of search experiments, so far,
together with astrophysical limits, constrain $F\gsim 10^9$ GeV.

Now let us consider the cosmological implications.   Peccei-Quinn symmetry is
unbroken at temperatures $T\gg F$.   When this symmetry breaks the initial value of
the  phase, that is
$e^{ia/F}$, is random beyond the then-current horizon scale.  One can analyze the
fate of these fluctuations by solving the equations for a scalar field in an expanding
Universe.   

The main general results are as follows.  There is an effective cosmic viscosity, which
keeps the field frozen so long as the Hubble parameter $H \equiv \overdot R/R \gg  m$, where $R$ is the expansion factor.  
 In the opposite limit
$H \ll m$ the field undergoes lightly damped oscillations,  which result in an energy
density that decays as $\rho \propto 1/R^3$.  Which is to say, a comoving volume
contains a fixed mass.   The field can be regarded as a gas of nonrelativistic particles
(in a coherent state).   There is some additional damping at intermediate stages.    
Roughly speaking we may say that the axion field, or any scalar field in a classical
regime, behaves as an effective cosmological term for $H \gg m$ and as cold dark
matter for $H\ll m$.  Inhomogeneous perturbations are frozen in
while their length-scale exceeds
$1/H$, the scale of the apparent horizon, then get damped.   

\begin{figure}[hbt]
\negsp
\centerline{\BoxedEPSF{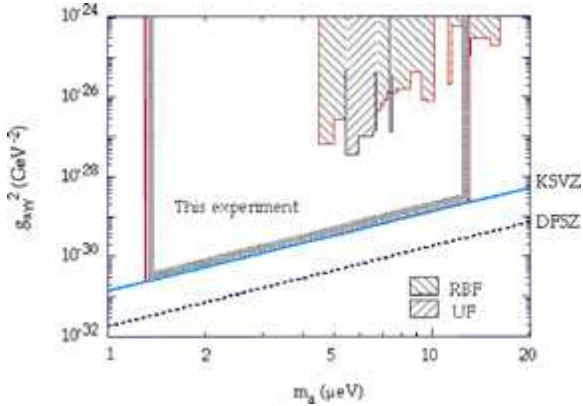 scaled 1500}}
\negsp
\caption{Experiments to search for a cosmic axion background.  They are approaching
sensitivities such that the hypothesis that axions contribute significantly to the dark energy will
be tested.}
\label{FWf18}
\negsp 
\end{figure}

If we ignore the possibility of inflation, then there is a unique result for the cosmic
axion density, given the microscopic model.    The criterion $H \lsim m$ is satisfied
for 
$T\sim \sqrt {M_{\rm Planck}/ F} \Lambda_{\rm QCD}$.   At this point the
horizon-volume contains many horizon-volumes from the Peccei-Quinn scale, but it
is still very small, and contains only a negligible amount of energy, by current
cosmological standards.    Thus in comparing to current observations, it is
appropriate to average over the starting amplitude $a/F$ statistically.    The result
of this calculation is usually quoted in the form 
$\rho_{\rm axion} / \rho_{\rm critical} \approx F/(10^{12}~{\rm GeV})$, where
$\rho_{\rm critical}$ is the critical density to make a spatially flat  Universe, which
is also very nearly the actual density.  Ongoing, heroic experiments are
approaching a crucial test of this dark matter candidate, as shown in
Figure~\ref{FWf18}.   

In the derivation of this form the
measured value of the baryon-to-photon ratio density at  present has been used.  This is
adequate for comparing to reality, but is inappropriate for our coming theoretical exercise.   If
we don't fix the baryon-to-photon ratio, but instead demand spatial flatness,  as
suggested by inflation, then what happens for 
$F > 10^{12}$ GeV  is that the baryon density is smaller than what we observe.   

If inflation occurs before the Peccei-Quinn transition, this analysis remains valid.  
But if inflation occurs after the transition, things are quite different.

\subsection{Undetermined Universe and \\the Anthropic Principle}

If inflation occurs after the transition, then the patches where $a$ is
approximately homogeneous get magnified to enormous size.  Each one is far larger
than the presently observable Universe.   The observable Universe no longer
contains a fair statistical sample of
$a/F$, but some particular ``accidental'' value.  Of course there is a still larger
structure, which Martin Rees calls the Multiverse, over which it varies.  

Now if $F>10^{12}$ GeV, we could yet be consistent with cosmological constraints
on the axion density, so long as the amplitude satisfies
$ (a/F )^2 \lsim F/( 10^{12}~{\rm GeV})$.   The actual value of $a/F$, which
controls a crucial regularity of the observable Universe, is contingent in a very
strong sense -- in fact, it is different ``elsewhere''.    In this very precise and specific sense, then, the laws
of physics are not unique.  

Within this scenario, the anthropic principle is correct and appropriate.  Regions
with large values of $a/F$, so that axions by far dominate baryons, seem pretty
clearly to be inhospitable for the development of complex structures.   The axions
themselves are weakly interacting and essentially dissipationless, and they dilute the
baryons, so that these too stay dispersed.  In principle laboratory experiments could
discover axions with $F > 10^{12}$ GeV.  Then we would conclude that the vast bulk
of the Multiverse was inhospitable to intelligent life, and we would be forced to
appeal to the anthropic principle to understand the anomalously modest axion
density in our Universe.  

\subsection{Coupling Nonuniqueness? -- \\ The Cosmological Term}

Ratcheting up the level of speculation one notch further, we can consider the
hypothesis that this is the {\it only\/} source of the observed non-vanishing
cosmological term.  To avoid confusion, let me call the axion variant that appears
here the {\it vadron}, in homage to Darth Vader (Lord of the Dark Force).   
I'll use the symbols $v$, $F_v$, etc. with the obvious
meaning.  

Several attractive consequences follow.  

\begin{itemize}
\item{The magnitude of the residual cosmological term is again of the general form
${\frac12} (v/F_v)^2 e^{{-8\pi^2}/{g_v^2}} \Lambda_v^4$ for
$v/F_v \ll 1$, then saturating, but now with
$g_v$ and $\Lambda_v$ no longer tied to QCD.   This could fit the observed value, for example, with
$v/F_v \sim 1$, $\Lambda_v \sim M_{\rm Planck}$, and 
$\alpha_v \sim 0.01$.}
\item{The freezing criterion $H \gsim m$ translates into $F_v \gsim M_{\rm
Planck}$.  If this condition holds by a wide margin, then the value of the effective
cosmological term will remain stuck on a time-scale of order $27H^{-1} (H/m)^4$,
considerably longer than the current lifetime of the Universe.  If $F_v$ is
comparable to or less than $M_{\rm Planck}$, significant conversion of the effective
cosmological term controlled by $v$ into matter is  occurring presently.}
\item{In any case, such conversion will occur eventually.   Thus we might be able to
maintain the possibility that a fundamental explanation will fix the asymptotic value
of the cosmological term at zero.}
\item{With larger values of $\alpha_v$ and smaller values of $v/F_v$, we realize an anthropic scenario, as discussed above, but now
for dark energy instead of dark matter.}
\end{itemize}

\subsection{Experiments!}

Speculation is fun (and I hope you'll forgive my indulgence) but to me the interest of an idea about Nature reaches a different level when its
truth can be tested by observation.   Lacking that discipline, we are in grave danger of saying things that are not merely wrong, but strictly meaningless
(Hume).   Fortunately, the idea of an ultralight sector does suggest several different kinds of concrete experiments.  I have already
mentioned the axion search.  There is also the possibility of searching for new quasi-macroscopic forces, both spin-independent (scalar
coupling) and spin-dependent (pseudoscalar coupling), or ``dipolar'', involving CP-violating cross-terms.   Another possible signature is
the time-variation of physical constants, which might be mimicked by interaction with a slowly varying cosmic background field. 
Pursuit of such heterodox experiments is difficult and risky, in the sense that negative results are not unlikely, but this pursuit is of
absolutely fundamental importance.   

\section{The Glory of Precision}

The most profound guiding principle of physics is that it is possible to construct a
precise quantitative description of basic physical phenomena.  (Nowadays it is PC to
add, what should go without saying, that in complex situations completely precise
description may not be practical.)   The special mission of high-energy physics --
perhaps its defining characteristic, more than high energy as such --  is to put this
principle to the test ruthlessly, at the most extreme limits of accurate calculation
and controlled measurement that we can attain.   It's both our glory  and  our burden
that we can care passionately about the existence of particles and effects that
emerge only from one-in-a-trillion collisions at ultra-sophisticated,
ultra-expensive accelerators, and worry over part-in-a-trillion discrepancies
between ultra-refined theory and ultra-delicate experiments.   

Two contemporary examples must be mentioned here.
\begin{figure}[hbt]
\negsp
\centerline{\BoxedEPSF{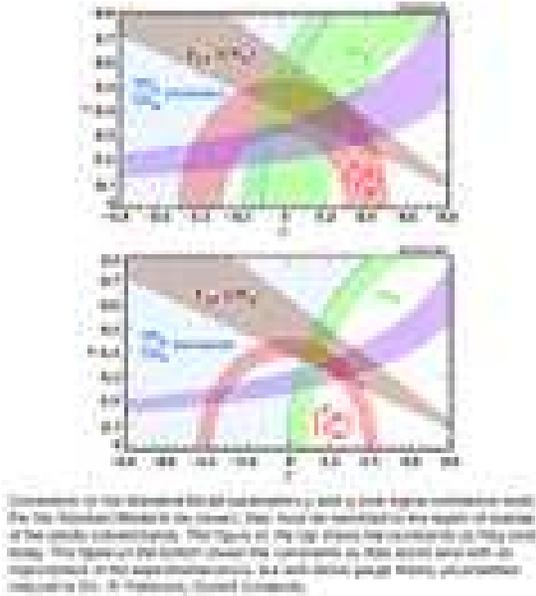 scaled 3400}}
\negsp
\caption{The unitarity triangle, again, together with an indication of how the power of the
measurements could be enhanced by achievable improvements in the theoretical estimation of
operator matrix elements in QCD.}
\label{FWf19}
\negsp  
\end{figure}

Figure~\ref{FWf19} illustrates how the value of experimental work on CP violation
and flavor physics could be enhanced by using the full power of QCD in its 
nonperturbative realization to calculate precise predictions for weak interaction
matrix elements.   It is not an isolated case.  There is a general principle at work:
Even if your ultimate ambition is to find deviations from the standard model, it's good
strategy, both scientifically and economically, to figure out as precisely as possible
what it is that the standard model predicts.  

We've just been treated \cite{semertzidis} to a description of new experimental results of
unprecedented accuracy for the anomalous magnetic moment of the muon.  Here
is the number: 
$$
a_\mu \equiv (g-2)/2 = 11~659~204~(7)(5)\times 10^{-10}.
$$  
It is an extraordinary achievement in experimental physics, the
fruit of remarkable courage, perseverance, and ingenuity.  Hats off!  With another
year of running time the precision, which currently is limited by statistics, could
be improved by a factor of two or more.  

We should not let memory of some recent pratfalls besmirch, even for a moment,
another magnificent story on the theory side.   Nowhere else in all of science are
such intricate calculations, involving concepts so remote from mundane
experience, both possible and necessary to do justice to experimental results.   Five
loops of QED!  Two loops of electroweak processes!   Dispersion relations, precise
low-energy experiments, and clever use of chiral symmetry for the virtual strong
interaction effects!   The result of all this is at once impressive, frustrating, and
tantalizing.   Depending upon whether one estimates hadronic vacuum
polarization based on $e^+e^-$ annihilation or $\tau$ decay, the discrepancy
between theory and experiment is 3.0 $\sigma$ or 1.9 $\sigma$.   Presumably the
difference between these estimates, which cannot represent real physics, will be
straightened out in coming months, possibly with an assist from lattice gauge
theory.   With achievable improvements in experiment and theory, conceivably
we could find ourselves looking at a 5 $\sigma$ effect by the next Lepton-Photon
get-together.  

It's not irrelevant to note that the anomalous moment of the muon has long been
recognized as an especially sensitive diagnostic for low-energy supersymmetry.   
Of course, by itself even a convincing discrepancy between experiment and the
standard model prediction of $g-2$ cannot rule low-energy supersymmetry in, nor
would the absence of such a discrepancy rule it out.  But if and when low-energy
supersymmetry is discovered, the value of $g-2$ will provide a very significant
piece of information concerning how the new sector communicates with ordinary
matter.

As we speak there are serious questions as to whether there will be adequate
funding available to exploit the existing opportunities, both in numerical QCD and
in experimental $g-2$.  In my opinion, failure to exploit these opportunities would be
a tragic waste.  We must strive harder, as a community, to convey the glory of
precision.  

\section{No Conclusion}

Breathe easy, there won't be a summary of the third order.   I would, however,
like to add one final observation.   The overwhelming impression I take away from
this Conference is that brilliant work by many people has resulted in an
extraordinarily precise, profound description of the physical world.  This
description incorporates on the one hand the vast body of knowledge and
technique quite inadequately labelled the standard model, enlarged to
accommodate neutrino masses; and on the other hand big bang cosmology, enlarged to
include the circle of ideas and observations around inflation.  But all this progress
should not mark an end.  Rather it allows us to ask -- that's easy enough! --  and
(more impressive) to take meaningful, concrete stabs at  {\it  answering\/} some
truly awesome questions.    Do all the fundamental interactions derive from a
single underlying principle?  What is the quantum symmetry of space-time? To
what extent are the laws of physics uniquely determined?  Why is there any
(baryonic) matter at all?  What makes the dark matter?   Why is there so little
dark energy, compared to what it ``should'' be?  Why is there so much, compared to
everything else in the Universe?   These are not merely popularizations or
vulgarizations but genuine, if schematic, descriptions of a few of our ongoing
explorations.

\subsubsection*{Acknowledgments} This work is supported in part by funds
provided by the U.S. Department of Energy (D.O.E.) under cooperative research
agreement
\#DF-FC02-94ER40818.

\end{document}